\pdfoutput=1 

\documentclass[5p]{elsarticle}

\usepackage{graphicx}
\usepackage{booktabs}
\usepackage{tabulary}
\usepackage{adjustbox}
\usepackage[hyphens]{url}
\usepackage[table]{xcolor}    
\usepackage{setspace}
\newcommand{\nz}{\ensuremath{0\nu\beta\beta}}

\newcommand{\bb}{\ensuremath{\beta\beta}}
\newcommand{\Q}{\ensuremath{Q_{\beta\beta}}}
\title{Background Discrimination for Neutrinoless Double Beta Decay in Liquid Xenon Using Cherenkov Light}
\author[llnl]{Jason Philip Brodsky}
\author[llnl]{Samuele Sangiorgio}
\author[llnl]{Michael Heffner}
\author[llnl]{Tyana Stiegler}
\address[llnl]{Lawrence Livermore National Laboratory}
\begin{document}
\begin{abstract}
    Neutrinoless double beta decays in liquid xenon produce a significant amount of Cherenkov light, with a photon number and angular distribution that distinguishes these events from common backgrounds. A GEANT4 simulation was used to simulate Cherenkov photon production and measurement in a liquid xenon detector and a multilayer perceptron was used to analyze the resulting distributions to classify events based on their Cherenkov photons. Our results show that a modest improvement in the sensitivity of neutrinoless double beta decay searches is possible using this technique, but the kinematics of the neutrinoless double beta decay and electron scattering in liquid xenon substantially limit this approach.
    
\end{abstract}

\maketitle
\section{Introduction and Physics Motivation \label{sec:intro}}
Neutrinoless double-beta decay (\nz) searches probe lepton number violation and Majorana mass of the neutrino \cite{majoranaTeoriaSimmetricaElettrone2008}. These experiments face the challenge of looking for an exceedingly rare signal and so background reduction plays a critical role in the choice of material used as a detection medium.

Liquid xenon-136 has been chosen by the EXO-200 \cite{augerEXO200DetectorPart2012} and nEXO experiments \cite{nexocollaborationNEXOPreConceptualDesign2018} as a double beta decay (\bb) source and detection medium due to a combination of properties including the ability to instrument liquid xenon as a time projection chamber resulting in good position and energy resolution.  In addition to the properties already used in liquid xenon experiments, the production of Cherenkov light in liquid xenon can enhance the ability of these detectors to differentiate neutrinoless double beta decays (\nz) from gamma backgrounds. This Cherenkov-based discrimination, while not currently a baseline feature of any experiment, could be implemented to reduce backgrounds and increase sensitivity in \nz\ searches.

\begin{figure}[tbp]
        \centering
        \includegraphics[width=0.9\columnwidth]{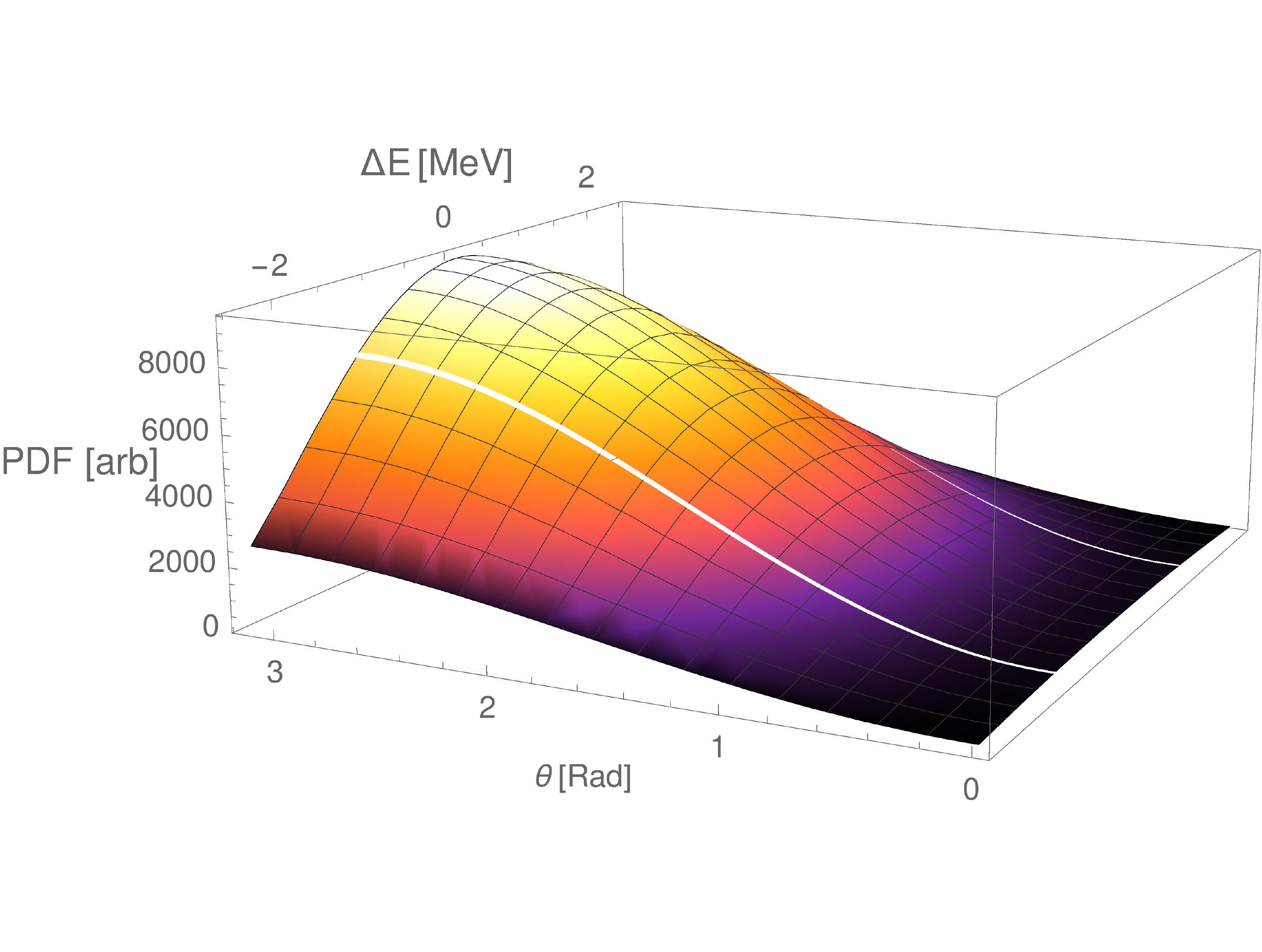}
        \caption{Distribution of neutrinoless double beta decays in energy split and relative angle. There are singularities at $\Delta E = \pm\Q/2$}
        \label{fig:neutrinodist}
        \end{figure}

The \nz\ process emits two electrons, which are both capable of producing Cherenkov light. These electrons have a total kinetic energy equal to the double beta decay Q-value (\Q\ = 2458 keV for $^{136}$Xe \cite{redshawMassDoubleBetaDecayValue2007}). The kinematics allows $\Delta E$, the difference in the electron energies, to fall anywhere in a wide distribution, although the "even split" of $\Delta E=0$ is the mode of that distribution \cite{boehmPhysicsMassiveNeutrinos1992}. Similarly, the angle between the two electrons' initial momenta is drawn from a wide distribution peaked at 180 degrees. Figure \ref{fig:neutrinodist} illustrates this distribution. As a result of this wide phase space, there is a great variety in the amount and angular distribution of the Cherenkov light emitted from \nz.
        
The primary background in neutrinoless double beta decay searches in liquid xenon arise from gamma rays with energy close to \Q. Some of these gamma rays undergo Compton scattering in the detector, producing multiple detected interaction sites that can be used to reject that event. However, gamma rays that undergo photoelectric absorption without any preceding scattering will produce a single recoiling electron with nearly all of the gamma's energy. Since electrons have short (few mm) range in liquid xenon, a single recoil electron background results in a single detectable interaction site just as the neutrinoless double beta decay does. For this reason, a site multiplicity cut cannot distinguish between this background and the signal. Further background reduction can be achieved by discrimination based on the spatial and energy distribution of backgrounds which have characteristic differences from the \nz\ signal, but even with all these techniques combined there remains a significant opportunity to improve the sensitivity of liquid xenon \nz\ detectors with further background reduction \cite{nexocollaborationSensitivityDiscoveryPotential2018}.

The background rate could be further reduced, and the sensitivity improved, by leveraging the differences in the Cherenkov photons produced between background and \nz\  events. The number and directional distribution of Cherenkov photons depends on the type of event, allowing for background discrimination. This article shows that in some scenarios this technique could remove more than 80\% of backgrounds while keeping more than 60\% of signal events, leading to \nz\ sensitivity improvements of about 40\%. We will show that these results rely on extremely efficient removal of Compton-scattering backgrounds which will otherwise prevent useful Cherenkov discrimination. We will examine some less favorable scenarios that may arise due to constraints on detector design. Finally, we will show that the barriers to more efficient Cherenkov discrimination are irreducible as they arise from the physics of \nz\ decay and electron scattering in liquid xenon.

This article starts by discussing the characteristics and detection of Cherenkov light in liquid xenon. Next comes analysis of a simulated test case demonstrating the degree to which differences in Cherenkov light production enable discrimination between \nz\  and photoelectric-absorbed gamma-ray backgrounds. Conclusions are drawn from a study of various physical and experimental factors that affect the discrimination power. 

\section{Cherenkov Production in Liquid Xenon\label{sec:chermodel}}
\begin{figure}[tbp]
    \centering
    \includegraphics[width=0.9\columnwidth]{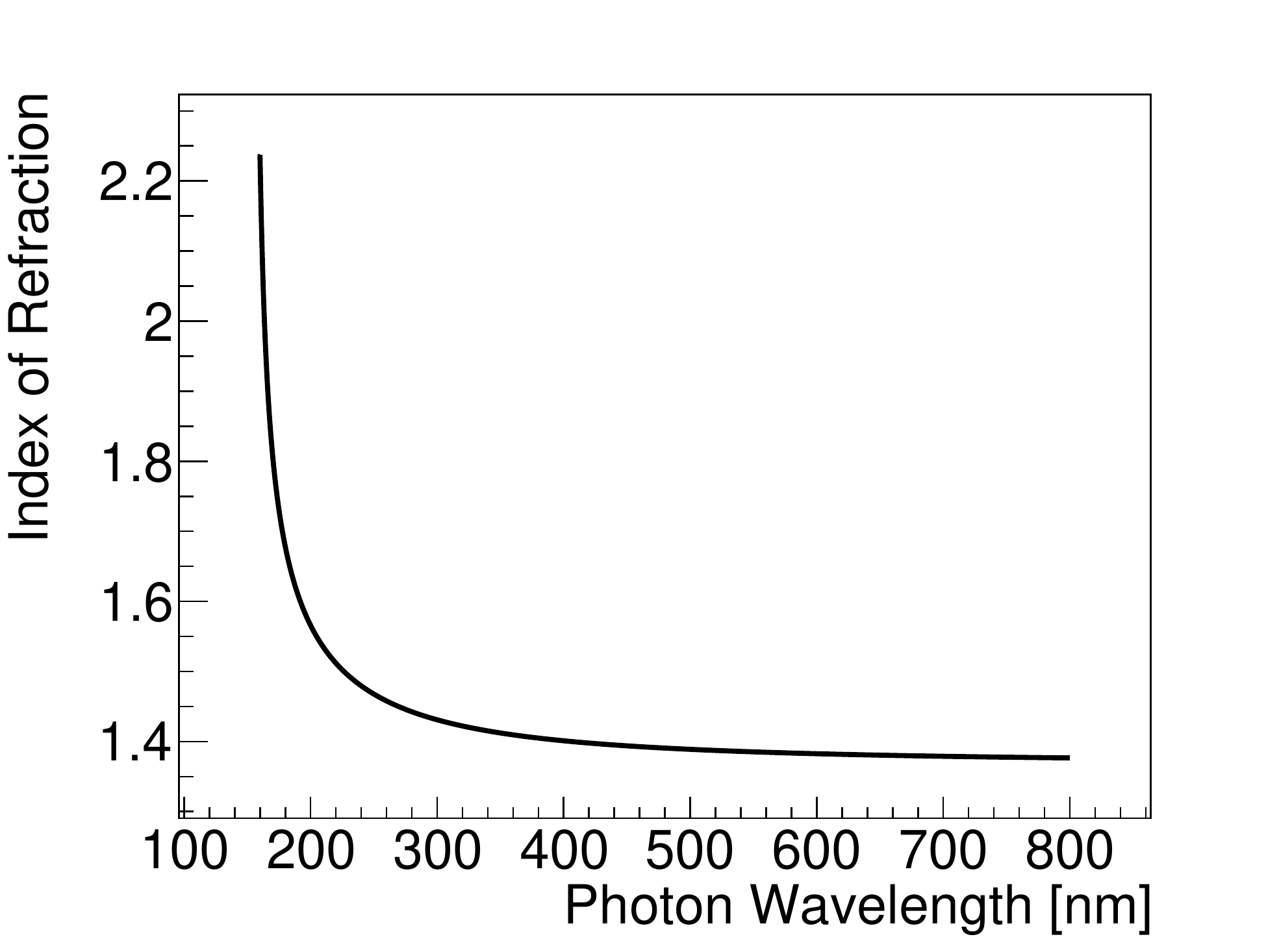}
    \caption{Liquid xenon index of refraction as a function of wavelength. Adapted from \cite{hitachiNewApproachCalculation2005}.}
    \label{fig:index}
    \end{figure}

    Cherenkov photons are produced when a charge particle, such as an electron emitted from a \nz, exceeds the phase velocity of light in the medium. As such, the number and spectrum of Cherenkov photons produced in liquid xenon depends on its index of refraction, which is wavelength-dependent.
    A functional form of the index of refraction of liquid xenon that agrees well with experimental measurements is described in \cite{hitachiNewApproachCalculation2005}. This model is used in this paper to simulate the production of Cherenkov photons and the transport of those photons through a detector. One input to this model is the number density of Xenon atoms, which was adjusted to correspond to 3.057 g/cc to account for the increased molar mass of xenon enriched in $^{136}$Xe \cite{nexocollaborationNEXOPreConceptualDesign2018}. The index of refraction as a function of wavelength is shown for reference in Figure \ref{fig:index}.
    
\begin{figure}[tbp]
    \centering
    \includegraphics[width=0.9\columnwidth]{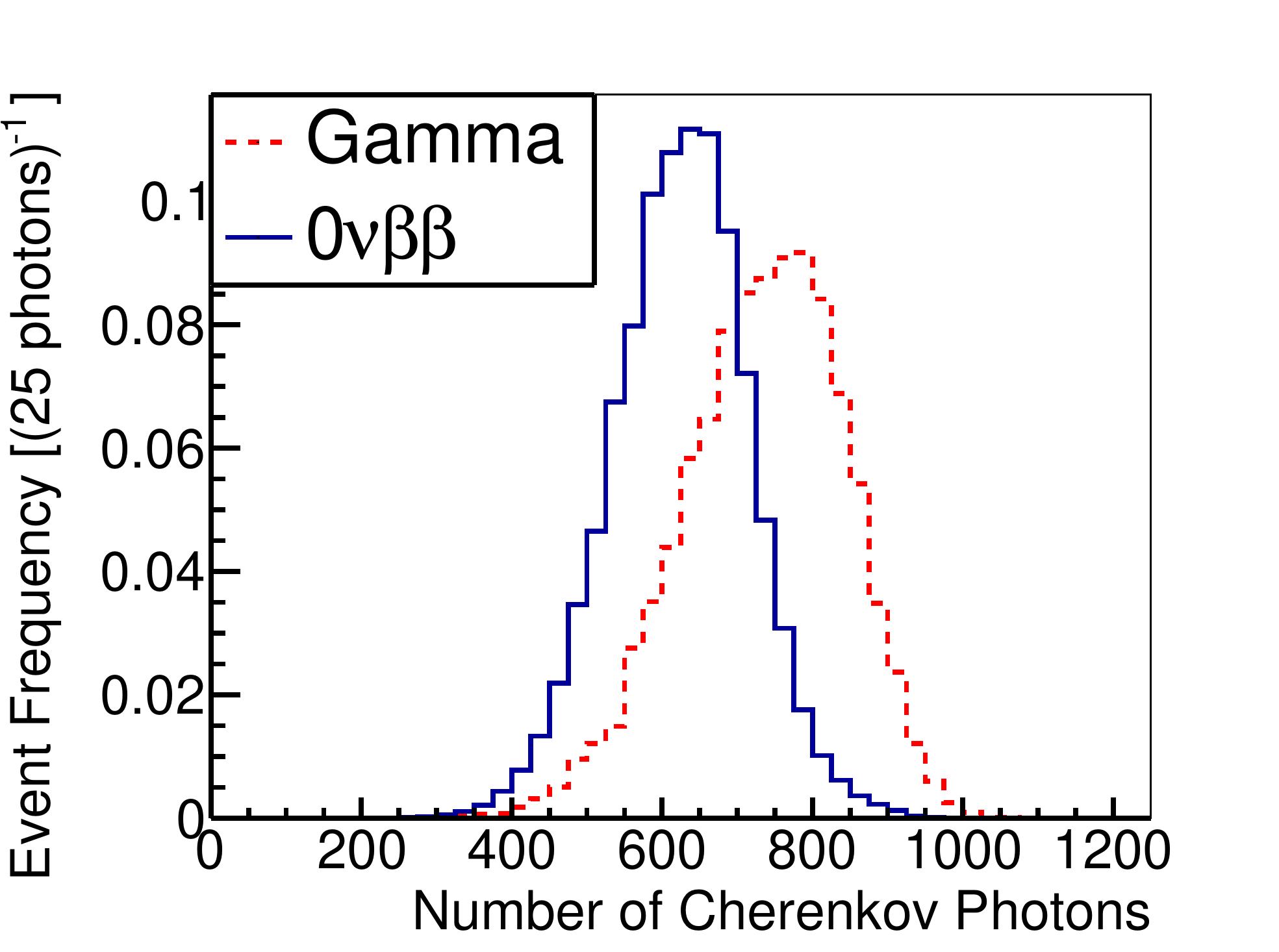}
    \caption{Cherenkov photon yield for $^{136}$Xe \nz\ and photoelectric interactions of gammas of \Q\ energy as obtained from the GEANT4 simulations used in this work.}
    \label{fig:nPhot}
    \end{figure}

    The Cherenkov yield from an electron can be estimated under the continuous-slowing-down approximation using stopping power values available from \cite{supleeStoppingPowerRangeTables2009}. Under this approximation, a single electron at \Q\  produces an average of 700 Cherenkov photons between 155 nm and 800 nm, which we treat as the sensitive range of a photosensor chosen for the vacuum-UV xenon scintillation light. The number of Cherenkov photons is nonlinear in energy, and so the yield for two electrons each with half of \Q\ is 582 photons in this approximation. When using the realistic distribution of energy split between the two electrons in \nz\ events\cite{boehmPhysicsMassiveNeutrinos1992}, the average yield is somewhat higher than the evenly-split decay case. Figure \ref{fig:nPhot} shows the Cherenkov yield in LXe obtained from a GEANT4\cite{agostinelliGeant4SimulationToolkit2003} simulation. The mean yield from the simulation of photoelectric interactions agrees approximately (within 5\%) with the continuous-slowing-down approximation for a single electron, but there is significant variation as each simulated electron slows down differently.
    
\begin{figure}[tbp]
    \centering
    \includegraphics[width=0.9\columnwidth]{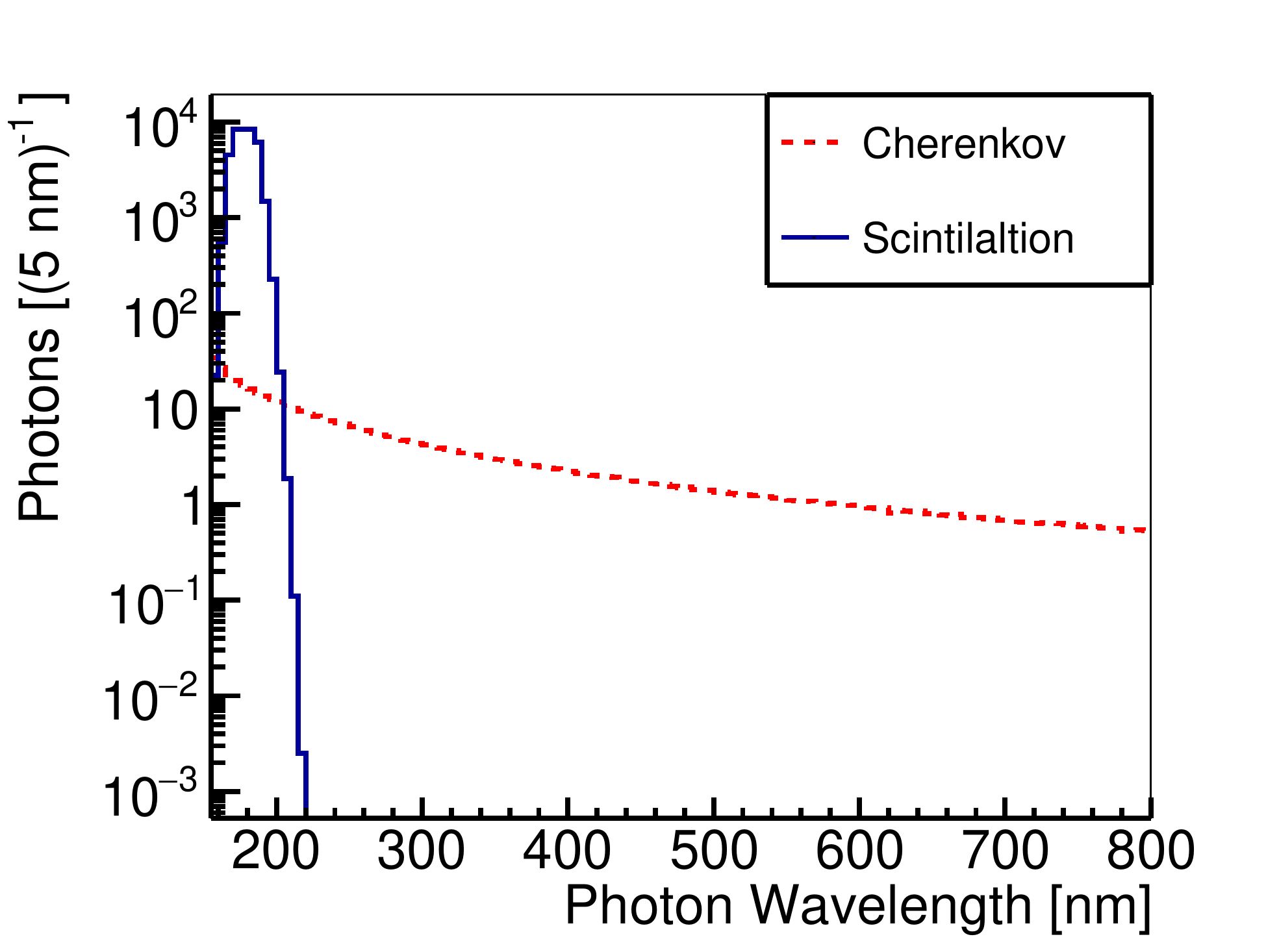}
    \caption{Simulated Cherenkov photon spectrum from \nz\ interactions in LXe compared to the corresponding scintillation spectrum.}
    \label{fig:typespectrum}
    \end{figure}

    Cherenkov photons are emitted in a wide distribution of wavelengths weighted heavily towards shorter wavelengths where the index is higher. Unlike Cherenkov light, scintillation light in liquid xenon is produced primarily in a narrow peak at 175 nm \cite{fujiiHighaccuracyMeasurementEmission2015}. The yield of scintillation photons for an electron at \Q\  is much higher that for Cherenkov photons, at an average of 89,500, as estimated based on \cite{lenardoGlobalAnalysisLight2015}. As a result, Cherenkov emission near the scintillation wavelength is impossible to measure, as illustrated in Figure \ref{fig:typespectrum}. 
    
    Background discrimination using Cherenkov light must rely on the photons at longer wavelengths than the scintillation light. These photons can be selected using a sensor insensitive to 175 nm light, or alternatively with a time-of-flight based cut. Here, we discard the former approach because it would compromise the detector's collection of scintillation light and therefore the precision of energy measurements. Precise energy measurements play a role in rejecting backgrounds in \nz\ searches and so a time-of-flight cut, which does not impact the energy measurement, is preferable.
    
\begin{figure}[tbp]
        \centering
        \includegraphics[width=0.9\columnwidth]{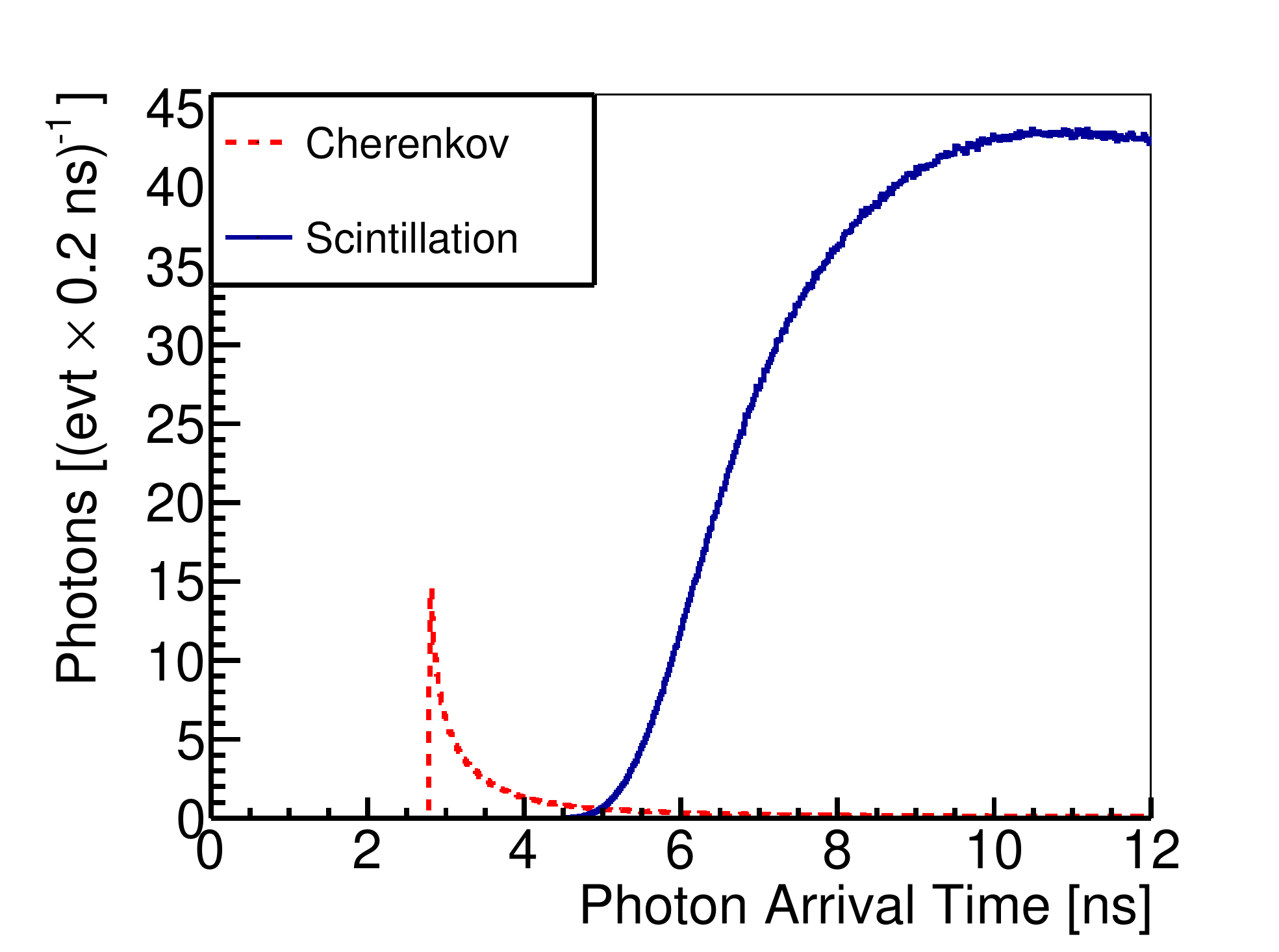}
        \caption{Distribution of arrival times for Cherenkov and scintillation light travelling a distance of 60 cm, obtained from simulated \nz\ interactions in LXe.}
        \label{fig:arrivaltime}
    \end{figure}

    A time-of-flight cut leverages the fact that Cherenkov photons outside the UV range travel substantially faster through liquid xenon than the UV scintillation light. The steep slope of the index of refraction near the scintillation wavelength results in a group velocity of scintillation light of $0.27 c$, compared to $0.42 c$-$0.71 c$ for Cherenkov light between 200 nm - 800 nm. In a detector with a 60 cm radius, this difference in speed results in the scintillation light from an event at the center of the detector arriving 3-4 nanoseconds later than the Cherenkov light as seen in Figure \ref{fig:arrivaltime}. This enables measurement of a portion of the Cherenkov light by looking only at those first few nanoseconds.
    
    In the following, we assume that the LXe scintillation light is completely negligible at wavelengths above 200 nm as it would otherwise  be difficult to distinguish Cherenkov light from scintillation with a time-of-flight cut or with any other method. Xenon is known to scintillate weakly in the infrared at $>800$ nm \cite{bressiInfraredScintillationComparison2001}, but there is negligible Cherenkov emission in that range. Argon does have a ``third continuum'' in its emission spectrum that produces some light between the UV and IR \cite{wieserNovelPathwaysAssignment2000} and no study has completely ruled out this mechanism in xenon. However, a study on xenon-doped argon \cite{neumeierIntenseVacuumUltraviolet2015} shows no scintillation between the UV and IR. Assuming the same holds true for pure liquid xenon, there would be no scintillation in a range that would compromise identification of Cherenkov photons by wavelength or time-of-flight. 
    
\begin{figure}[tbp]
        \centering
        \includegraphics[width=0.9\columnwidth]{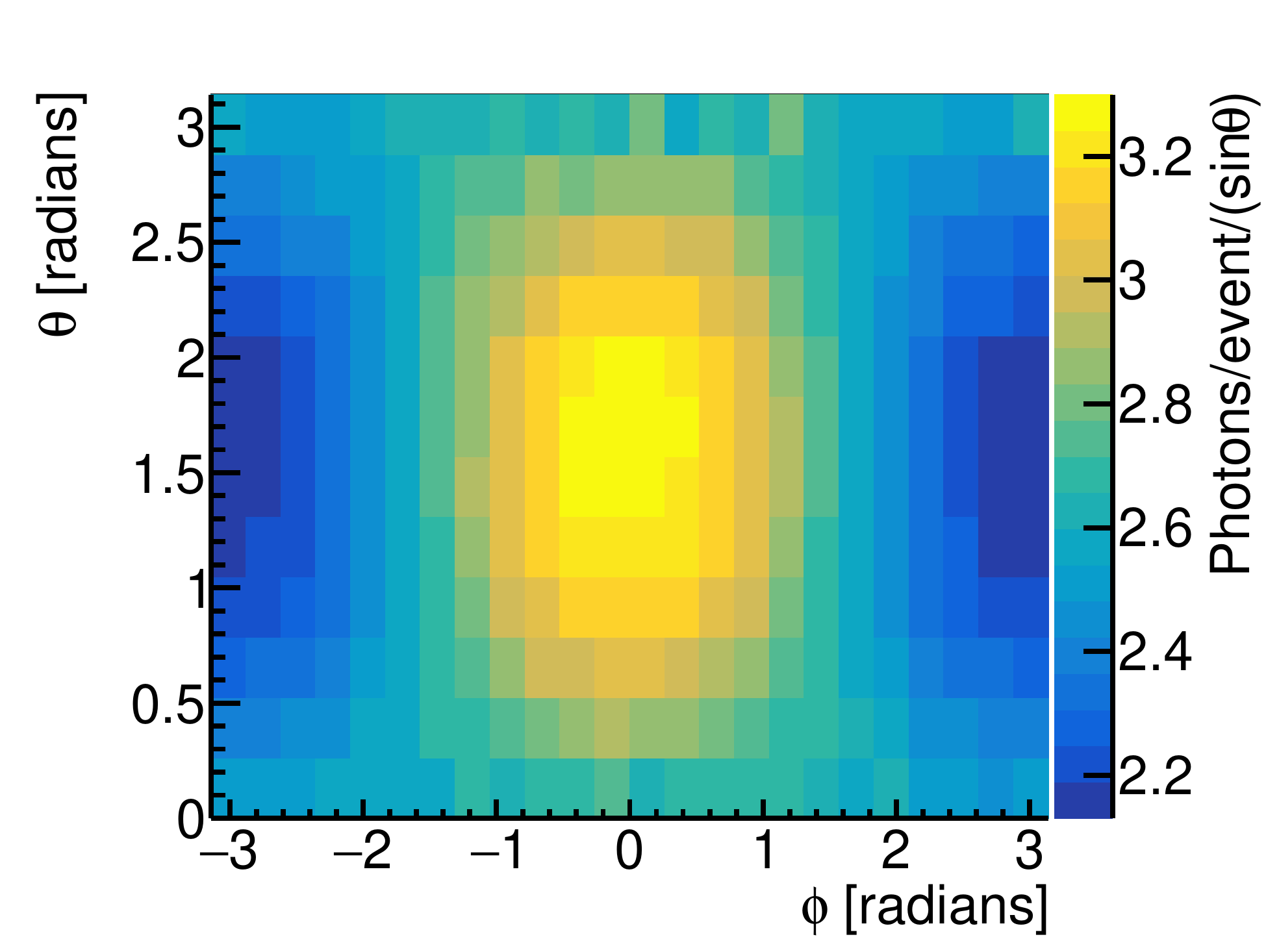}
        \caption{Direction of individual Cherenkov photons relative to the mean direction of Cherenkov photons in each simulated interaction of an electron with initial energy \Q. While no characteristic ring shape is observed, some directional correlation is preserved. }
        \label{fig:directionality}
    \end{figure}

    Cherenkov emission forms a characteristic cone relative to the electron's velocity, but MeV-scale electrons in liquid xenon do not produce the ring patterns on the light sensors familiar from water Cherenkov detectors measuring GeV-scale particles. In fact, electrons scatter frequently and at large angle in liquid xenon, so the emission angle of Cherenkov light is measured from a constantly-changing electron direction. Figure \ref{fig:directionality} shows the direction of emission of simulated Cherenkov photons from electrons with \Q\ initial energy. The direction of each individual Cherenkov photon is plotted relative to the average direction of photons from that event. The observable directional bias shows that the Cherenkov photon directions are correlated, not isotropic, and so retain some information about the direction of charged particles in the event. 
    
    Rayleigh scattering of scintillation and Cherenkov photons adds random noise to the relationship between electron direction and photon direction. This scattering also increases the mean and variance of the distance travelled by photons between the event site and the light sensors. The Rayleigh scattering length depends greatly on the photon wavelength, being 35 cm at the scintillation wavelength and rising to 1 m by 200 nm photons and 75 m by 436 nm photons. For purposes of this study, Rayleigh scattering is neglected for photons with wavelengths greater than 436 nm as in those wavelengths the scattering length is much larger than the detector sizes simulated here.


\section{Simulation and Analysis Methodology}

    A GEANT4 \cite{agostinelliGeant4SimulationToolkit2003} simulation was performed to determine if neutrinoless double beta decay could be distinguished from backgrounds using Cherenkov light in a variety of scenarios. The results were analyzed to quantify the amount of discrimination possible and to project the impact of that discrimination on the sensitivity of a \nz\ search.
    
    The simulation and the subsequent analysis do not represent any existing or proposed detector for several reasons. No existing or proposed detector has been designed to take best advantage of Cherenkov signals, and so the baseline capabilities of those detectors are not informative for a study of what might be accomplished in a detector optimized for that purpose. For the sake of easily exploring different scenarios affecting the power of Cherenkov-based discrimination, we used several simplifying assumptions. When choosing parameters affecting the simulation or analysis, we chose the most favorable scenario within reason as a baseline and investigated some alternative scenarios with less favorable assumptions. The results can therefore be taken as an upper limit of what might be achieved, with some illustration of how far below that upper limit a realistic detector might fall. However, we made every effort to ensure the underlying physics, as described in section \ref{sec:chermodel}, is accurately represented in the simulation. The specific simplifications and favorable assumptions are highlighted in the remainder of this section. 
    
    GEANT4.10.3 was used for the simulation. The Livermore electromagnetic model \cite{allisonRecentDevelopmentsGeant42016} governed the behavior of electrons and gammas in the xenon. Cherenkov production was handled using the GEANT4 Cherenkov process with default parameters. Scintillation photons were produced using a modified version of the NEST tool, described in Section \ref{sec:NEST}.
    
    The baseline detector geometry  was a sphere of xenon with 60 cm radius, with the entire surface of the sphere made of photodetectors assumed to be 30\% efficient at detecting all light reaching this surface. This spherical arrangement was chosen to simplify the analysis of the simulation by taking advantage of the symmetries of the geometry. It also forms a best-case scenario, as it has complete photodetector coverage and the light is not absorbed or scattered by internal hardware between the xenon and the photosensors, such as electric field-shaping hardware. This baseline design contains 2.7 tonnes of liquid xenon, and so is referred to as the ton-scale sphere design.
    
    Several alternative geometries were simulated in addition to the baseline. A kiloton-scale sphere design was one alternative geometry, with a 428 cm radius and containing 1 kilotonne of liquid xenon. Two ton-scale cylinder designs were simulated, both with 60 cm radius and 120 cm height and containing 4.1 tonnes of xenon. These two designs differed in their photosensor coverage. The first had the entire surface, cylinder wall and bases, instrumented with photosensors, similar to the spherical designs. The second had only the wall instrumented, while the bases were uninstrumented but 80\% reflective at all wavelengths.


    \subsection{Simulated Neutrinoless Double Beta Decays\label{sec:simsignal}}
        \nz\ events were simulated as the production of two electrons whose energies and directions were determined by the kinematics of the double beta decay process. In the GEANT4 simulation, these decays were produced by a custom generator code that drew the beta energies and directions from the appropriate distributions as described in section \ref{sec:intro}.
        
        Neutrinoless double beta decays were simulated only at a single point at the center of the detector volume. This reduced the complexity of the analysis, avoiding the need to account for event position. Events at the center of the detector, where external gamma backgrounds are the lowest, also have the most impact on the limit-setting or discovery potential of a liquid xenon neutrinoless double beta decay experiment \cite{nexocollaborationSensitivityDiscoveryPotential2018}. Thus, events at this location are the most interesting for tests of the ability to reject external gamma backgrounds. 
        
        The events in this study have an inherent symmetry due to being located at the detector center. This symmetry is used to reduce the number of simulated events needed in the training set by rotating the coordinate system as described below in Section \ref{sec:analysis_details}. Studies at points without this symmetry will require a larger population of simulated events but will not otherwise be hampered by the lack of symmetry.
        
        Simulated \nz\ events were also required to pass a single-interaction-site cut. The typical \nz\ event produces a single continuous energy deposition. Some \nz\ events, however, produce bremsstrahlung gammas that take energy further from the main event location. When this occurs, the event resembles a gamma background with separate Compton and photoelectric interactions. Following the method in \cite{nexocollaborationSensitivityDiscoveryPotential2018}, we reject from the analysis \nz\ events with energy depositions more than 3 mm separated from the main energy cluster at the center of the detector.

    \subsection{Simulated backgrounds \label{sec:backgrounds}}
    
        We simulated gamma rays of energy \Q\ biased to interact in the center of the detector, as in the signal case. We furthermore required the backgrounds to pass a single-interaction-site cut. We examined two different versions of this cut, one perfectly efficient at identifying single interaction sites and another based on the cut described in \cite{nexocollaborationSensitivityDiscoveryPotential2018} which includes inefficiency from imperfect position resolution.
        
        The perfectly efficienct single-site cut requires that all simulated background gammas immediately undergo photoelectric effect at the center of the detector. This is implemented using GEANT4's event biasing options, producing a set of simulated events with only a single interaction site where the photoelectric effect occurred. Bremsstrahlung gammas emitted from the photoelectric-effect electron may travel far enough from the electron's energy deposition track that they produce a second detectable interaction site, in which case we reject the background event as non-single-sited. We consider this to have occurred and reject the background event if a Bremsstrahlung gamma travels more than 3 mm from the electron's track. The perfectly efficient single-site cut is the baseline for analysis.
        
        An experiment's single-site cut may be less than perfectly efficient due to limitations on resolving the event topology. In particular, a gamma that undergoes Compton scatter and then photoelectric effect within a very short distance may be impossible to distinguish from a pure photoelectric effect event.  Reference \cite{nexocollaborationSensitivityDiscoveryPotential2018} describes a single-interaction-site cut that includes some reasonable assumptions about experimental resolution, and we implemented a cut following that method. Energy deposits within 3 mm of each other were aggregated into clusters. If more than one such cluster is found, i.e. if there is an energy deposit more than 3 mm from an existing cluster, the background event is discarded. The remaining events include both gammas that immediately underwent photoelectric effect interaction, producing a single recoiling electron, and events that underwent one or more Compton scatters within a region of about 3 mm diameter before undergoing photoelectric effect, producing multiple recoiling electrons that are aggregated into a single cluster. 43\% of the events were in photoelectric-only category. The other 67\% of events represent an inefficiency in the single-site cut, as they include multiple interaction sites that are too close to distinguish. 
        
        We do not use this imperfect cut as the baseline because, as shown later in section \ref{sec:results}, this inefficiency is highly detrimental to Cherenkov-based discrimination, as multiple recoiling electrons from a background are difficult to distinguish from the two electrons in a \nz\ event.
        This inefficiency may not be easily reducible due to the experimental realities of electron diffusion during drift \cite{nexocollaborationNEXOPreConceptualDesign2018}, but it is possible that future development will result in a cut that more efficiently removes the Compton-scattering backgrounds. As the next generation of liquid xenon \nz\ detection refines its projections of the efficiency of the single-site cut, we will learn to what degree Cherenkov-based discrimination can approach the performance seen in this purely-photoelectric simulation. For the time being, we use the more favorable scenario as a baseline to more easily explore the implications of Cherenkov-based discrimination.

    \subsection{Light Simulation\label{sec:NEST}}
        Optical photons were generated from two separate Geant4 processes. Scintillation was generated using a modified version of the Noble Element Scintillation Technique (NEST) tool for Geant4. These modifications increased the computing performance of NEST when working with MeV-scale events. This version of NEST produces a yield of 36.4 photons/keV for a Q-value electron recoil in a 300 V/cm electric field. 
        
        The G4Cerenkov process controlled Cherenkov photon production. 
        Geant4's implementation of the calculations behind Cherenkov production is particularly sensitive to the smoothness of the input curve relating index of refraction to photon wavelength, so we supplied a curve with 10,000 points between 155 nm and 800 nm, with the index calculated at each point using the model described in Section \ref{sec:chermodel}.
        
        Photons were propagated using Geant4 from their origin to the edge of the detector, where their location and wavelength were recorded. This simulation process does not model non-light-sensitive hardware in the active volume. A 30\% efficiency was assumed for detection of photons incident on the light sensing surface of the xenon vessel. This represents the photon detection efficiency of the sensors used, distinct from the light collection efficiency that includes absorption of light elsewhere in the detector.
        
        The simulated detector does not include internal hardware that would either absorb or reflect light before it encounters the light sensing surface, reducing the collection efficiency or delaying light collection past the time-of-flight cut. This hardware is commonly present in detectors, but a detailed simulation of its effects is beyond the scope of this work. Instead, we considered a scenario treating the light sensing efficiency as 10\% to demonstrate the effect of photons lost to absorption or reflections on hardware. Even the 10\% light sensing scenario is more favorable than currently existing detectors which lack the total light sensing coverage in the baseline simulated detector. Absorption and scattering within the liquid xenon is included in all simulations.
    
    \subsection{Event Analysis\label{sec:analysis_details}}
        
        First, the photons collected in the simulation were filtered through a time-of-arrival cut that separated scintillation photons from the Cherenkov photons. For each simulated geometry, the location of detection for each photon was divided by the time of arrival to get an implied speed of the photon. The cut threshold was tuned manually to find the value that maximized the Cherenkov discrimination ability. This cut was $0.427 c$ for the ton-scale geometries and $0.379 c$ for the kiloton-scale geometries, which required a different cut due to the change in photon transport caused by Rayleigh scattering over long distances. 41\% of Cherenkov photons fail this cut in the ton-scale geometry, as either their wavelength is closer to the scintillation wavelength or they reflect or scatter on their way to detection, increasing path length to longer than the distance between the origin and the detection and reducing the implied speed below the cut. In the kiloton-scale geometry, that increases to 57\%.
        
        The simulation results were processed to produce, for each event, a map of the number of photons striking each location in the detector. Position information was binned into 24-by-12 "pixels" by polar coordinates, so that each event can be summarized by a vector of photon quantities of length equal to the number of pixels. The number of pixels was chosen by starting with a small number of pixels and increasing the number until the performance of the analysis plateaued. For consistency in the analysis, polar coordinates were used even when the simulated geometry was cylindrical. The simulated detectors are symmetric around one (cylinder) or two (sphere) axes, so two events with similar Cherenkov photon distributions apart from the rotation should be analyzed the same way. To that end, during the binning of the photon results the coordinate system was rotated event-by-event. For spherical geometries, the coordinate system pointed the primary axis along the dipole moment of the photon distribution and the secondary axis in an arbitrary direction. For cylindrical geometries, the primary axis always pointed along the axis of symmetry while the secondary axis lay in the same plane as the dipole moment of the photon distribution. The effect of these rotations was that the greatest concentration of photons in each event would consistently appear in the same bins, simplifying the last step of the analysis.
        
        The signal-background discrimination analysis was performed by training a multi-layer perceptron (MLP) to classify events as either signal or background. The MLP technique was implemented using the Python package Scikit-learn \cite{pedregosaScikitlearnMachineLearning2011} which based its approach on \cite{hintonConnectionistLearningProcedures1990} and \cite{kingmaAdamMethodStochastic2014}. We chose a machine learning approach for this problem because it could flexibly learn the best way to apply directional information in the distribution of Cherenkov photons. MLPs are a well-tested choice of machine learning algorithm for analysis of images used in event classification \cite{kolanoskiApplicationArtificialNeural1996}. Some otherwise-appropriate alternatives, such as support vector machines, proved impractical due to the speed at which their computational requirements scale with the large data set used here.
        
        Six million events were provided to the MLP, of which 600,000 were reserved for testing, and the remainder were used as the training set.  A range of hyperparameters were tried, as well as different non-MLP algorithms to conclude that the hyperparameters given in Table \ref{tab:hyperparams} produced sensitivity results, as evaluated in Section \ref{sec:sensitivity} as good or better than any other option. The hidden layers used were fully connected, as Scikit-Learn does not support other architectures.
        
        \begin{table}[tbp]
            \centering
            \begin{tabular}{c|c}
                Hyperparameter & Value \\
                \hline
                Hidden layers & 100, 100 \\
                Activation function & Hyperbolic tangent \\
                Optimization algorithm & Adam \cite{kingmaAdamMethodStochastic2014} \\
                Learning rate $\alpha$ & $10^{-3}$ \\
                Batch size & 50
            \end{tabular}
            \caption{Multilayer Perceptron Hyperparameters}
            \label{tab:hyperparams}
        \end{table}
    
    \subsection{Sensitivity Analysis \label{sec:sensitivity}}
        The goal of neutrinoless double beta decay experimental design is to maximize the sensitivity of the experiment to the half-life of neutrinoless double beta decay. We use a simplified model of experimental sensitivity to estimate the improvement in sensitivity that can be achieved from Cherenkov discrimination.
        
        Although real experiments typically use a more sophisticated analysis to determine their sensitivity \cite{nexocollaborationSensitivityDiscoveryPotential2018}, in this simplified model we assume that events are simply counted and compared to the expected number of backgrounds. 
        The Feldman-Cousins sensitivity \cite{feldmanUnifiedApproachClassical1998}, i.e. the average upper limit of experiments under the null hypothesis, is then be calculated. 
        This simplification is necessary as the sensitivity estimates in \cite{nexocollaborationSensitivityDiscoveryPotential2018} are closely tied to the distribution of backgrounds throughout the detector volume while this study of Cherenkov-based discrimination examines only the center of the detector.
        
        The Feldman-Cousins sensitivity is in units of signal counts and can be converted to a limit on the neutrinoless double beta decay half-life by a factor proportional to the signal acceptance. So, if $FC(b)$ is the Feldman-Cousins sensitivity of a counting experiment with $b$ backgrounds, $f$ is the fraction of backgrounds passing the Cherenkov discrimination cut, and $s$ is the signal acceptance of that cut, the half-life sensitivity improvement $I$ is:
        
        \begin{equation}
        \label{eq:sens}
            I = s \cdot \frac{FC(b)}{FC(b\cdot f)}
        \end{equation}

        The half-life sensitivity improvement depends on an assumption of $b$, the number of backgrounds before Cherenkov discrimination is applied. Each experiment has its own value of $b$ that depends on the experimental design. For smaller values of $b$, the sensitivity impact of Cherenkov discrimination is smaller, a consequence of the behavior of the Feldman-Cousins sensitivity when expected backgrounds are close to zero. As a baseline, we assume $b=100$. This baseline case represents the application of Cherenkov-based discrimination by an experiment with a relatively large background rate to achieve a competitive number of backgrounds. We also examine a scenario in which $b=10$, which represents the application of Cherenkov-based discrimination in an experiment that has already achieved low backgrounds and seeks to use this technique for further reduction.

\section{Simulation Results \label{sec:results}}

    As mentioned above, the baseline scenario is a ton-scale spherical geometry using 24-by-12 pixels in the analysis, with backgrounds consisting of only photoelectric-effect interactions. In this baseline scenario, an average of 133 photons from each background event pass the Cherenkov cuts, compared to 109 photons from each signal event.
        
\begin{figure}[tbp]
    \centering
    \includegraphics[width=0.9\columnwidth]{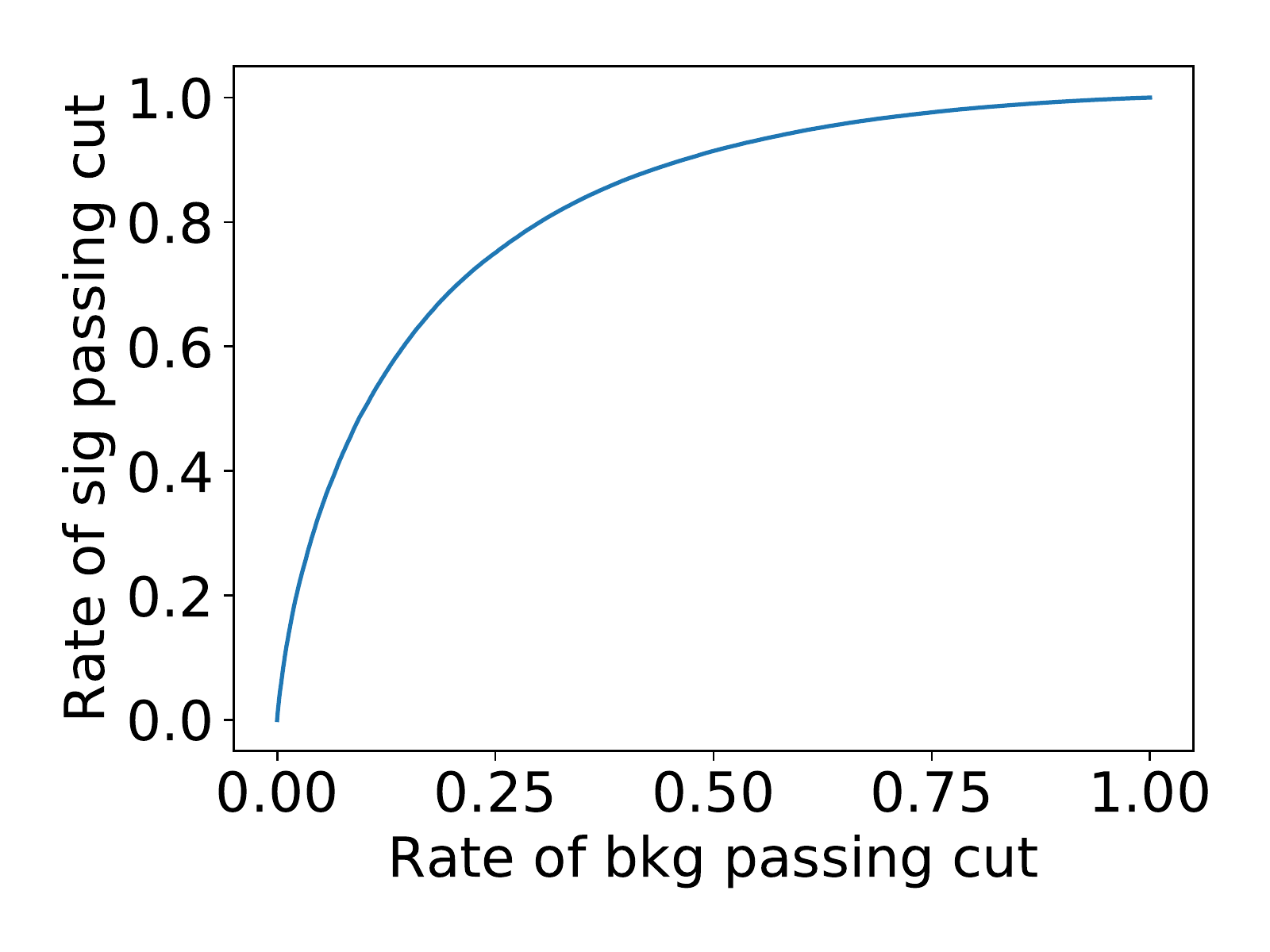}
    \caption{ROC curve of ton-scale sphere with baseline analysis}
    \label{fig:m8_roc}
    \end{figure}

    Applying the MLP analysis produces the Receiver Operating Characteristic (ROC) curve in figure \ref{fig:m8_roc}. Each point on this ROC curve is a potential cut threshold on the MLP metric with the indicated consequence on background rejection vs signal acceptance.
    
\begin{figure}[tbp]
    \centering
    \includegraphics[width=0.9\columnwidth]{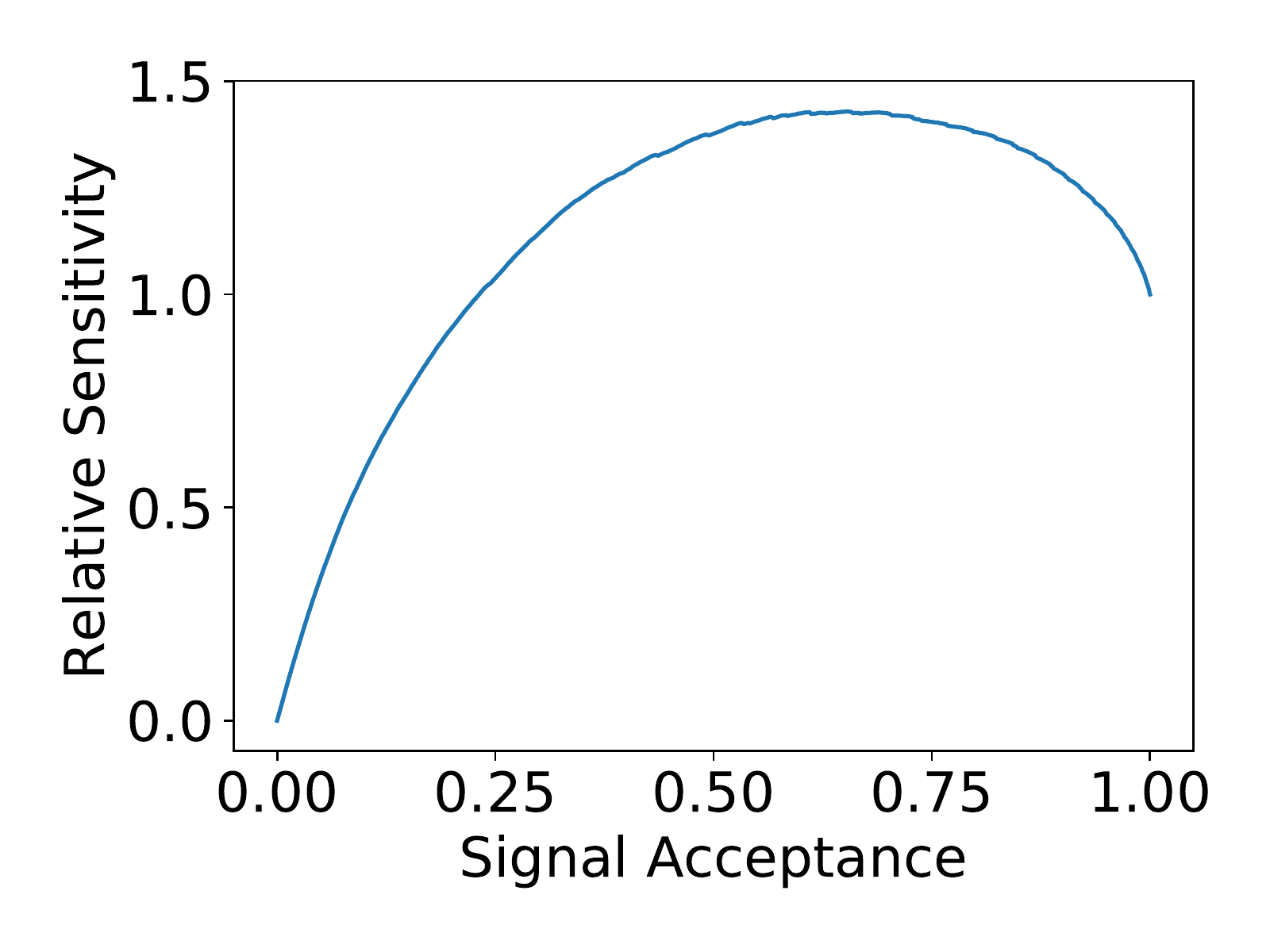}
    \caption{Sensitivity as a function of background rejection in a Ton-Scale Sphere with Baseline Analysis}
    \label{fig:m8_sens}
    \end{figure}

    Using the methodology in Section \ref{sec:sensitivity}, the ROC curve in Figure \ref{fig:m8_roc} can be translated into the sensitivity improvement curve in Figure \ref{fig:m8_sens}. The maximum of this curve is the best improvement in the sensitivity that can be achieved using a cut on the MLP metric. In this baseline case, the best sensitivity improvement is a factor of 1.43 improvement at a cut that removes 85\% of backgrounds and keeps 61\% of signals.
    
    These promising results highly depend on the single-sited cut used to create the background set. The less efficient cut described in section \ref{sec:backgrounds} includes many Compton-scattering backgrounds. Performing the Cherenkov analysis using this background set produces only a factor of 1.11 sensitivity improvement. This number was produced assuming $b=200$ in the sensitivity calculation, to represent the increase in the pre-Cherenkov-discrimination backgrounds from the less efficient single-site cut. The origin of this reduced discrimination effectiveness is clear: when a gamma produces multiple recoiling electrons due to Compton scatter, the resulting Cherenkov light is difficult to distinguish from \nz\ events.
    
    We simulated several other scenarios to explore how effective Cherenkov discrimination is under different assumptions. Table \ref{tab:results} summarizes the results of these other scenarios. It provides for each case short description and a number, which can be referenced to the more detailed description below. It then provides the results for each case: the sensitivity improvement at the optimal cut value, the number of Cherenkov photons passing cuts in the background and \nz\ test sets, and the signal acceptance  and background rejection that enables the reported sensitivity improvement.
    
        \rowcolors{2}{gray!25}{white}
        \begin{table*}[tbp]
        \centering
        \tymin 20 pt
        \caption{Result of the study of Cherenkov \nz\ background discrimination in liquid xenon for various scenarios. In each case, the sensitivity improvement factor $I$ is provided along with the number of Cherenkov photons per signal (\nz) and background event, the signal acceptance, and the background rejection obtained from the MLP analysis. See text for detailed description of each scenario.  } \label{tab:results}

        \begin{tabulary}{\textwidth}{@{}L R R R R R R@{}}

Case	& Description	& Sensitivity improvement	& Signal {photons/event} & Background photons/event	& Signal acceptance	&	Background rejection \\
\toprule
1	&	Baseline	&	1.43	&	109	&	133	&	0.61	&	0.85	\\
2	&	Compton Scatters included	&	1.11	&	109	&	121	&	0.80	&	0.49	\\
3	&	Perfect background rejection	&	7.61	&	N/A	&	N/A	&	1.00	&	1.00	\\
4	&	Back-to-back evenly-split \nz\ 	&	1.96	&	36	&	133	&	0.68	&	0.91	\\
5	&	Back-to-back, even split \nz\ and straighter tracks	&	5.53	&	143	&	111	&	0.79	&	1.00	\\
6	&	Truth-value Cherenkov ID	&	1.40	&	183	&	218	&	0.68	&	0.80	\\
7	&	100\% detection efficiency	&	1.59	&	362	&	442	&	0.64	&	0.87	\\
8	&	10\% detection efficiency	&	1.20	&	36	&	44	&	0.75	&	0.64	\\
9	&	No directional information	&	1.34	&	109	&	133	&	0.75	&	0.72	\\
10	&	Kiloton sphere	&	1.35	&	76	&	93	&	0.66	&	0.80	\\
11	&	Kiloton sphere w/o direction	&	1.30	&	76	&	93	&	0.73	&	0.72	\\
12	&	Cylindrical geometry, 4pi instrumentation	&	1.43	&	108	&	131	&	0.65	&	0.83	\\
13	&	Cylindrical geometry, side-only instrumentation	&	1.38	&	87	&	106	&	0.63	&	0.82	\\
14	&	Sensitivity assuming b=10	&	1.22	&	133	&	109	&	0.79	&	0.71	\\
15	&	Perfect background rejection, b=10	&	2.76	&	N/A	&	N/A	&	1.00	&	1.00	\\

        \bottomrule
        \end{tabulary}
        \end{table*}

    \subsection{Understanding the Limitations of Cherenkov Discrimination}
    
    The baseline scenario produces relatively modest improvements in sensitivity. To investigate why the improvements are not greater, we simulated deliberately unphysical scenarios in which Cherenkov-based discrimination may be more effective.
    
    First, we evaluated the sensitivity of an idealized experiment that manages to perfectly reject all backgrounds with no loss of signal (Table \ref{tab:results}, Case 3). This result comes directly from equation \ref{eq:sens}. The sensitivity improvement factor of 7.61 represents an upper bound on what any background rejection technique can achieve in a scenario where b=100.
    
    Next, we replaced the realistic simulation of \nz\  with a simplified model in which the betas are always emitted precisely back-to-back and with exactly half the Q-value energy each (Case 4). In this physically unrealistic scenario, the sensitivity improvement was a factor of 1.96. This scenario aids the discrimination because it creates the most contrast between a single photoelectric electron and a background event. The directional information of two back-to-back electrons is as different as possible from the single-electron background events. The overall Cherenkov photon yield of two \Q/2 electrons is less than that from any other way to divide \Q\  between the two electrons or the yield of a single electron background. This scenario was a significant improvement over the baseline case, suggesting that handling the wide double beta decay kinematic phase space is a major challenge for Cherenkov discrimination.
    
    The gap between the idealized \nz\ Case 4 and the perfect discrimination Case 3 can be largely explained by the tendency of electron tracks in liquid xenon to scatter frequently. We applied a cut on the simulated backgrounds and idealized \nz\ used in Case 4, keeping only events in which the primary recoiling electron(s) travel further from the event origin than the 90th percentile. By selecting the longest-travelling electrons, we choose a set of events with less-than-typical amounts of electron scattering to demonstrate the effect that scattering has. This scenario (Case 5) has excellent sensitivity improvement, a factor of 5.53, close to the performance of the perfect discrimination case. This demonstrates that electron scattering is the primary limiting factor in Cherenkov discrimination.
    
    We ran a scenario to evaluate the impact of the time-of-flight cut used to separate Cherenkov and scintillation photons (Case 6). All scintillation photons were removed from the analysis and all Cherenkov photons were included whether or not they passed the time-of-flight cut. The 30\% detection efficiency was still applied. The discrimination power in this case was somewhat below the baseline (1.40 vs 1.43), suggesting that losing some Cherenkov photons to the time-of-flight cut actualy improves the discrimination. We believe this is because the Cherenkov photons that would fail the time of flight cut do not carry valuable information about the nature of the event. These are largely photons from the shorter wavelength part of the Cherenkov spectrum which move more slowly through the xenon causing them to fail the time-of-flight cut. These photons are more likely to Rayleigh scatter and so contribute noise to the directional information. The relative difference between signal and background photon counts also goes down, as the number of short-wavelength photons depends less on electron energy than the number of longer-wavelength photons.
    
    Finally, we examined the influence of collection efficiency on the discrimination. The baseline scenario fixed a 30\% detection efficiency for Cherenkov photons. A simulated scenario with 100\% efficiency (case 7) produces a sensitivity improvement of 1.59. This is a smaller increase than case 3, suggesting that while capturing more Cherenkov photons would help additional photon statistics cannot overcome the limitations imposed by electron scattering and double beta decay kinematics. A simulation at 10\% efficiency (case 8) shows that low photon statistics do lead to a much smaller sensitivity improvement factor of only 1.20.
    
    \subsection{Potential Losses of Cherenkov Discrimination Power}
    
    The model of baseline scenario did not include some factors that may be present in some real detectors that will further reduce Cherenkov discrimination power. This section describes additional scenarios that were studied to estimate the impact these additional factors might have.
    
    Additional hardware, such as field shaping rings, may scatter Cherenkov light and reduce the available directional information. In the worst case, this would produce output in which the location where the Cherenkov light was collected carries no information. We modelled this worst case scenario using an analysis that replaced the 24-by-12 photon position binning with a single bin (Case 9). This removes all directional information from the analysis, resulting in reduced discrimination ability and a sensitivity improvement of 1.34. This confirms that the directional information in the Cherenkov light is useful in distinguishing signal from background, but the number of photons alone is sufficient to perform some discrimination.
    
    In the kiloton-scale geometry (Case 10), the increased distance between the center of the detector and light sensors at the detector edge causes photons to undergo more Rayleigh scattering during their flight. In addition to blurring the photon directional information, the Rayleigh scattering blurs the timing information, making the time-of-flight cut less effective at separating Cherenkov photons from scintillation photons. The time-of-flight cut threshold was optimized separately from the smaller detector scenario, but still includes significantly fewer Cherenkov photons: an average of 76 photons from \nz\ events in the kiloton-scale detector versus 109 in the ton-scale detector. These effects reduced the discrimination ability compared to the baseline case producing a sensitivity improvement of 1.35.
    
    Combining the above two cases, a kiloton-scale detector was analyzed using only one photon position bin (Case 11). The result here was a sensitivity improvement of 1.30, worse than using the same one-bin analysis on a smaller detector (Case 9). This confirms that the Raleigh scattering's effect on the time-of-flight cut is significant and the larger detector's performance loss is not purely due to the Rayleigh scattering obscuring directional information.
    
    To understand the effect of detector geometry a cylindrical geometry was evaluated. A spherical geometry was chosen for the baseline case to simplify the analysis, but cylindrical detector designs are common. The cylindrical design had the same radius as the spherical geometry, resulting in a larger mass of xenon. We studied two cases, one in which all interior surfaces of the cylinder could detect photons (Case 12) and another in which only the side of the cylinder while the bases of the cylinder were 80\% reflective but not instrumented (Case 13). The full-instrumentation cylinder had similar performance to the baseline spherical case, while the cylinder with non-instrumented bases had reduced performance, with a sensitivity improvement of 1.38.
    
    Finally, we considered the effect that our assumption of $b=100$ in section \ref{sec:sensitivity} had on our results by finding the maximal sensitivity improvement in the baseline scenario with $b=10$ (Case 14). Changing $b$ does not affect the ability of the Cherenkov discrimination to reject backgrounds and accept signals, but affects the benefit in sensitivity at any given rejection/acceptance and puts the optimum sensitivity at a different point on the ROC curve. Reducing $b$, i.e. assuming fewer backgrounds exist to be removed via Cherenkov analysis, reduces the sensitivity benefit of that analysis down to a factor of 1.22. For reference, in a scenario with perfect background rejection and signal acceptance and $b=10$ (Case 15) the sensitivity improvement is 2.76.

\section{Conclusion}

These results indicate the potential for Cherenkov-based discrimination in liquid xenon \nz\ detectors. A baseline case with optimistic assumptions allows for discrimination using time-of-flight-selected Cherenkov photons that preserves 61\% of the neutrinoless double beta decays while removing 85\% of photoelectric backgrounds. In a detector with 100 backgrounds in the experiment duration, this discrimination would improve sensitivity to the \nz half-life by a factor of 1.43. This result is highly dependent on the cuts applied to background before Cherenkov analysis is performed, and so any inefficiency in the single-interaction-site cut that removes Cherenkov-scattering backgrounds will strongly impact the potential for Cherenkov discrimination.

We have shown that intrinsic properties of Cherenkov production from \nz\ decay events limit the discrimination ability. The scattering of Cherenkov-emitting electrons in liquid xenon and the kinematics of the \nz\ decay are responsible for most of the difference between the achievable performance of Cherenkov-based discrimination and a perfect discrimination method. Detector features that reduce the light collected or increase photon scattering will also reduce the value of Cherenkov-based discrimination.

This result also depends on several simplifications used in the simulation and analysis. The most significant of these is that events were only simulated in the center of the detector. If the discrimination ability varies significantly outside the center of the detector the whole-detector sensitivity improvement result must be adjusted to reflect that. The extension of this work to events outside the detector center requires additional complexity in the event generation and analysis tools used and will be the focus of future work.

These results differ from those in \cite{PossibleUsageCherenkov2016}, which identified a greater separation between signal and background events. We attribute the difference in results primarily to a difference in the Cherenkov yield of background events between \cite{PossibleUsageCherenkov2016} and this work. Reference \cite{PossibleUsageCherenkov2016} reports 20\% fewer Cherenkov photons in \nz\ events than in background events, compared to a 15\% difference in yield in this work.
    
\section*{Acknowledgements}
This work was performed under the auspices of the U.S. Department of Energy by Lawrence Livermore National Laboratory under Contract DE-AC52-07NA27344 and was supported by the LLNL-LDRD Program under Project No. 18-ERD-028.

\section*{References}
\bibliography{MyLibrary} 
\end{document}